\documentclass[aps,prd,floats,amssymb,showpacs,nofootinbib,preprint]{revtex4}
\usepackage{graphicx}
\usepackage{amsmath}
\usepackage{amssymb}
\usepackage{epsfig}

\newcommand{\be}{\begin{equation}}
\newcommand{\ee}{\end{equation}}
\newcommand{\bea}{\begin{eqnarray}}
\newcommand{\eea}{\end{eqnarray}}
\newcommand{\nn}{\nonumber}
\newtheorem{prop}{Proposition}

\newcommand{\lag}{{\cal L}}
\newcommand{\D}{{\cal D}}
\newcommand{\h}{{\cal H}}
\newcommand{\M}{{\cal M}}
\newcommand{\N}{{\cal N}}
\newcommand{\R}{{\cal R}}
\newcommand{\ms}{M_*}
\newcommand{\mn}{\mu\nu}

\newcommand{\Sum}{\displaystyle\sum}
\newcommand{\Lim}{\displaystyle\lim}

\def\bphi{\bar\phi} 
\def\bchi{\bar\chi} 
\def\pphi{\varphi} 
\def\pchi{\xi} 
\def\ux{u} 

\begin{document}

\begin{flushright}
\mbox{\normalsize \rm UMD-PP-10-001}\\\vspace{-10pt}
\end{flushright}


\title{Scalar Kinks in Warped Extra Dimensions
}
\author{Manuel Toharia$^{a}\!\!$ \footnote{mtoharia@umd.edu}}
\author{Mark Trodden$^{b}\!\!$ \footnote{trodden@physics.upenn.edu}}
\author{Eric J. West$^{b,c}\!\!$ \footnote{ejwest@physics.syr.edu}}

\affiliation{
$^a$ Maryland Center for Fundamental Physics, Department of Physics, 
University of Maryland, College Park, MD 20742, USA
\\
$^b$ Center for Particle Cosmology, Department of Physics and Astronomy, 
University of Pennsylvania, Philadelphia PA 19104, USA 
\\
$^c$ Department of Physics, 
Syracuse University, Syracuse NY 13244, USA}

\date{\today}

\begin{abstract}
We study the existence and stability of static kink-like configurations of a 5D scalar field, with Dirichlet boundary conditions, along the extra dimension of a warped braneworld. In the presence of gravity such configurations fail to stabilize the size of the extra dimension, leading us to consider additional scalar fields
with the role of stabilization. We numerically identify multiple nontrivial solutions for a given 5D action, made possible by the nonlinear nature of the background equations, which we find is enhanced in the presence of gravity. Finally, we take a first step towards addressing the question of the stability of such configurations by deriving the full perturbative equations for the gravitationally coupled multi-field system. 
\end{abstract}

\maketitle


\section{Introduction}
\label{intro}

The possibility of extra spatial dimensions~\cite{Kaluza:tu,Klein:tv}, hidden from our current experiments and observations through compactification or warping, has opened up a wealth of options for particle physics model building~\cite{Rubakov:1983bb,Akama:1982jy,Antoniadis:1990ew,Lykken:1996fj,Arkani-Hamed:1998rs,Antoniadis:1998ig,Randall:1999ee,Randall:1999vf,Lykken:1999nb,Arkani-Hamed:1999hk,Antoniadis:1993jp,Dienes:1998vg,Kaloper:2000jb,Cremades:2002dh} and has allowed entirely new approaches for addressing cosmological problems~\cite{Arkani-Hamed:1998nn,Macesanu:2004gf,Starkman:2001xu,Starkman:2000dy,Deffayet:2001xs, Deffayet:2001pu,Dvali:2000hr,Deffayet:2002sp,Deffayet:2000uy,Binetruy:1999hy, Binetruy:2001tc,Binetruy:1999ut,Chung:1999zs,Csaki:1999mp,Cline:2002ht,Nasri:2002rx}. 
In many implementations, Standard Model (SM) fields can be confined to a submanifold, or brane, while in others they populate the entire extra-dimensional space. Common to both approaches, however, is the inclusion of bulk fields beyond pure gravity, either because they are demanded by a more complete theory, such as string theory, or because they are necessary to stabilize the extra-dimensional manifold. Thus, a complete understanding of the predictions and allowed phenomenology of extra dimension models necessarily includes a comprehensive consideration of the configurations of these bulk fields, the simplest of which are real scalars. Indeed, 4D Poincare invariance allows for these new bulk fields to acquire nontrivial static configurations along the extra dimensions. 

Static one-dimensional scalar configurations with a node (where the field vanishes) are known to localize wave functions of other fields near that node. In the context of extra dimensions, these kink-like scalar backgrounds can be used for example to localize bulk fermions near either boundary~\cite{Arkani-Hamed:1999dc, Georgi:2000wb,Kaplan:2001ga,Grzadkowski:2004mg}, allowing for interesting constructions of flavor models. They can also affect the localization of other scalar or vector fields leading to a field theoretic description of fat branes (see for example the constructions in \cite{Hung:2003cj,Surujon:2005ia,Davies:2007xr}). Kink-like scalar configurations are a particularly interesting case to consider because the boundary conditions make it possible to obtain non-trivial general results regarding both the existence and the stability of such configurations, at least in the case of one flat extra dimension without gravity~\cite{Toharia:2007xe,Toharia:2007xf}. In this paper we  build on these previous results and extend them as far as possible to the case with a gravitating (warped) extra dimension.  In the presence of gravity the kink-like configuration cannot fix and stabilize the interbrane distance~\cite{Lesgourgues:2003mi}. It is therefore necessary to assume the existence of at least one additional stabilizing field, coupling either directly or gravitationally to the kink field.  We will opt for the latter and introduce additional non-interacting scalar fields. At least one of these additional fields must be given a monotonic profile in order to stabilize the size of the extra dimension (i.e., interbrane modulus)~\cite{Goldberger:1999uk}. 

The plan of the paper is as follows. In section II we review the results for kink-like backgrounds in a flat extra dimension with no gravity. We then generalize these to include a warped gravitational background, but with no gravitational backreaction from the kink-scalar itself. In section IV we consider the coupled system of multiple scalar fields in the presence of gravity, and as a special case consider a kink field with Dirichlet boundary conditions and another scalar whose purpose is to stabilize the whole configuration. We write the background equations for such a system and show graphically how non-linearities allow a given action to have multiple static solutions.

In the final section we take the first steps towards studying the question of the stability of these static configurations by deriving the complete set of equations for scalar and gravitational perturbations around a given static background. The general procedure is quite complex and involves extended theorems of oscillation theory appropriate to the type of eigenvalue problem we are lead to, namely a matrix Sturm-Liouville problem. We therefore reserve a complete study of the stability of the general system for future work.


\section{Kinked Scalars in Flat Extra Dimensions}
\label{flatxd}

In~\cite{Toharia:2007xe,Toharia:2007xf} a 5D flat scenario including
one real scalar field with an arbitrary scalar potential was
studied and the general conditions for the existence and perturbative 
stability of static, nontrivial, background scalar field
configurations were presented. In this section we briefly review the
main results and slightly extend the discussion of the energy densities
of different kink configurations. 

Consider a real scalar field in 5 dimensions (labeled by indices $M,N,\ldots =0,1,2,3,5$) with a flat background metric, and defined by the action  
\be
   S = \int d^5x\,\left[
   \frac{1}{2} \eta^{MN} (\partial_M \phi) \;\partial_N \phi 
   - V(\phi)\right]\ .
   \label{action_flat_metric}
\ee
The extra dimension is compactified on an $S_1/Z_2$ orbifold with the
scalar field $\phi(x,y)$ being odd under $Z_2$ reflections along the
extra coordinate $x^5\!\equiv\!y$ (i.e. $\phi(x,y)=-\phi(x,-y)$). Here
the orbifold interval is defined as $[0,\pi R]$, with its size $\pi R$
assumed to be fixed. The potential $V(\phi)$ must then be invariant
under the discrete symmetry $\phi\to-\phi$, and is chosen to have at
least two degenerate global minima at $\phi\!=\!\pm v$, with
$v\!\neq\!0$. To simplify notation, we will also choose the potential
to vanish at $\phi\!=\!0$. 

\begin{figure}[t!]
  \center
  \includegraphics[width=16cm,height=8cm]{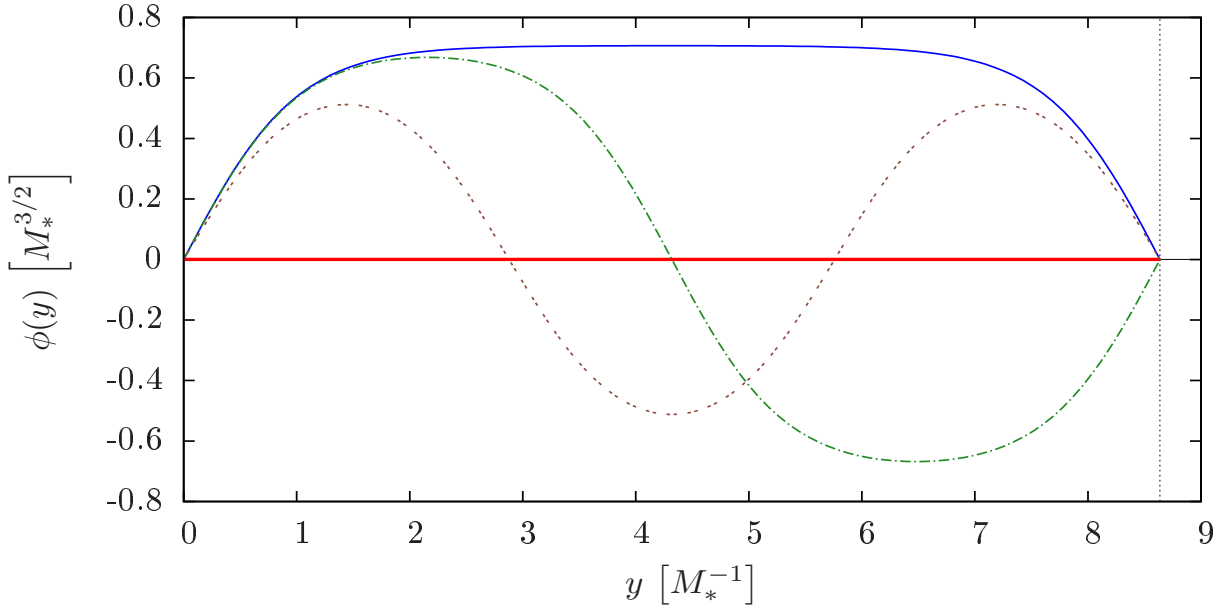}\ \
  \vspace{0.2cm}
  \caption{Profiles in the extra dimension interval $[0,\pi R]$ 
  of different static configurations of the Dirichlet scalar field 
  $\phi$, defined by the scalar potential 
  $V(\phi)=-\frac{1}{2}|\mu^2|\phi^2+|\lambda|\phi^4$
  ($\mu^2\!=\!2\ms^2$, $\lambda\!=\!1\ms^{-1}$, 
  $\pi R\!=\!\!8.6375\ms^{-1}$). The solutions with nodes in the interval
  (dashed curves) are unstable, while the stability of the 
  nodeless and trivial solutions (solid curves) depend on the parameters
  of the model.}
  \label{kinks}
\end{figure}

Under these conditions, it was shown in \cite{Toharia:2007xe} that
there will always be static solutions, nontrivial along the extra
coordinate $y$, satisfying the (static) field equation 
\be
   \phi'' - \frac{\partial V}{\partial \phi} = 0 \ ,
   \label{scal_el_general}
\ee
where a prime denotes a derivative with respect to $y$. The profiles
of these solutions, satisfying Dirichlet boundary conditions, resemble
that of a kink solution patched to an anti-kink in the middle of the
interval. The possible solutions were classified in two groups, namely
those with nodes in the interval (multiple kink-antikink solutions patched
together) and those with no nodes, vanishing only at the end-points of
the orbifold (see Fig.~\ref{kinks}).  It was shown that all static
kink solutions with nodes are perturbatively unstable, whereas the
stability of nodeless solutions depends on the parameters of the model
in a particularly simple way. 

\begin{figure}[t!]
  \center
  \includegraphics[width=16cm,height=8cm]{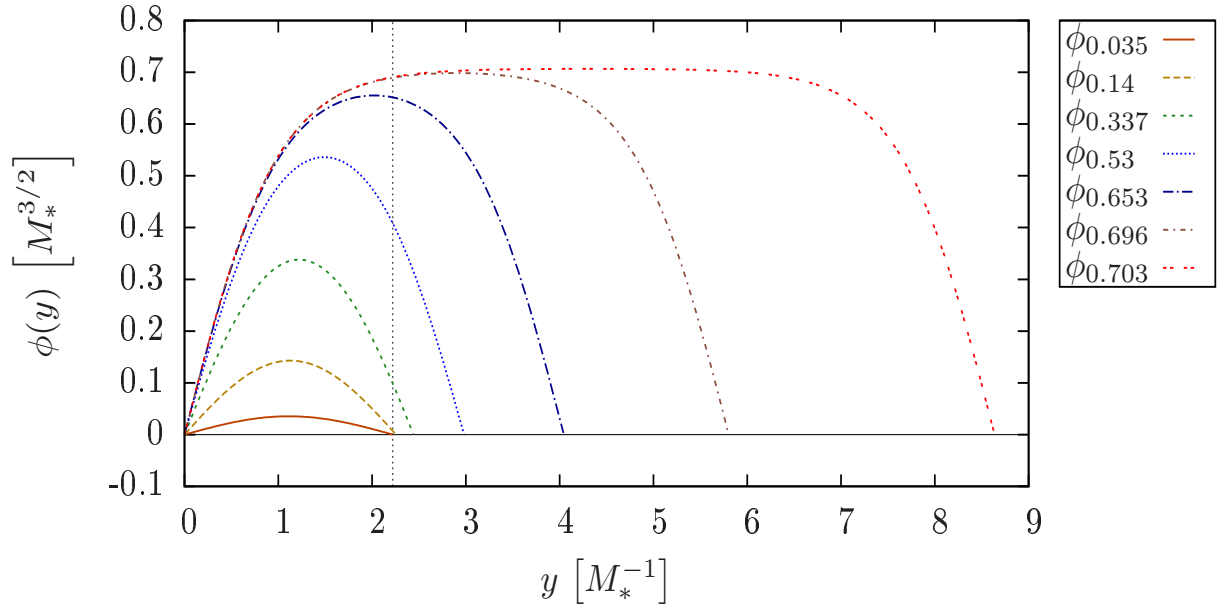}
  \vspace{0.2cm}
  \caption{Nodeless static configurations of the kink scalar field 
  $\phi$, defined by the scalar potential
  $V(\phi)=-\frac{1}{2}|\mu^2|\phi^2+|\lambda|\phi^4$ 
  ($\mu^2\!=\!2\ms^2$, $\lambda\!=\!1\ms^{-1}$). Configurations with 
  different amplitudes are solutions to different physical problems, 
  corresponding to different stabilization radii of the extra dimension. 
  The vertical dashed line indicates the minimal radius $R_c$, below which nodeless
  solutions do not exist with this potential.}
  \label{nodelesskinks}
\end{figure}

The Dirichlet solutions of Eq.~(\ref{scal_el_general}) with no nodes
in the interval form a continuous one-parameter family of functions. A
simple choice for the parameter is the amplitude $A$ of the solution,
i.e., the maximum value of the nontrivial solution
$\phi_A(y)$. Solutions with different amplitudes $A$ generally vanish
at different points along the extra dimension, which correspond to
different possible orbifold radii $R$ (see
Fig.~\ref{nodelesskinks}). However, in order to obtain the stability
condition for these solutions it is extremely useful to consider the
full family of solutions.  

The value of the 4D effective energy density of a
given static solution $\phi_A(y)$ is 
\be
   E(A) = \int_0^{T(A)} \left(\frac{1}{2}{\phi'_A}^2 +
   V(\phi_A)\right) dy \ , 
\ee
where $T(A)$ is the length of the solution in the extra
dimension. This can be conveniently rewritten as an integral over
$\phi$ using properties of Eq.~(\ref{scal_el_general}) and its
solutions $\phi_A(y)$ 
\be
   E(A) = 2\sqrt{2}\int_0^A \frac{V(A)-2\ V(\phi)}{\sqrt{V(\phi)-V(A)}}\,d\phi \ .
   \label{energyvsA}
\ee
We are now equipped to state the general results of
\cite{Toharia:2007xe,Toharia:2007xf} in a slightly modified, although
more revealing, version:
\begin{prop}
A static solution to equation~(\ref{scal_el_general}),
with $\delta >0$ nodes inside the orbifold interval is always unstable.
\end{prop}
\begin{prop}
A static, nodeless solution $\phi_{A_*}(y)$ to
equation~(\ref{scal_el_general}), with amplitude $A_*$, and associated
energy density $E(A_*)$ is stable if 
\be
   \left.\frac{dE}{dA}\right|_{A=A_*} < 0 \ .
   \label{Estabilitycriterion}
\ee
\end{prop}
\noindent This is a powerful result since it means that given any
scalar potential $V(\phi)$ we immediately know
which of the nontrivial nodeless solutions $\phi_A$ will be stable or
unstable, without the need to actually know explicitly their analytic
form.  

With this result it is possible to understand the vacuum structure of
any single scalar field theory with Dirichlet boundary conditions in 5D when the metric along
the extra dimension is flat. Possible static solutions consist of the
trivial solution $\langle\phi\rangle=0$ (which may or may not be stable), kink-like
solutions with nodes in the interval (which are always unstable), and
kink-like solutions without nodes in the interval (some stable and some
unstable, depending on condition (\ref{Estabilitycriterion})). As
remarked in \cite{Toharia:2007xe,Toharia:2007xf}, the trivial solution
may be the true vacuum solution even in the case of a negative mass
term $-|\mu^2| \phi^2$ in the 5D potential, as long as the inequality
$|\mu^2|<|1/R^2|$ is preserved. Therefore, for a given orbifold radius $R$, many different perturbatively stable
vacuum solutions are possible, and it is necessary to identify which
one is the true vacuum of the theory.  

The true vacuum of the theory will depend on the size of the radius $R$. This
can be seen as follows: Without loss of generality, one may define the
energy density of the trivial solution to be zero by choosing the 5D
potential $V(\phi)$ to vanish at $\phi=0$. It was shown
in~\cite{Toharia:2007xe,Toharia:2007xf} that there is a critical
radius $R_{c}$ below which nontrivial nodeless solutions do not exist
(see Fig.~\ref{nodelesskinks}). The energy density associated with the
critical nontrivial nodeless solution will 
be either positive or exactly zero, so that the transition from one
vacuum to another can be either second order or first order, as
one varies the radius $R$. 

\begin{figure}[ht!]
  \center
  \includegraphics[width=16cm,height=6cm]{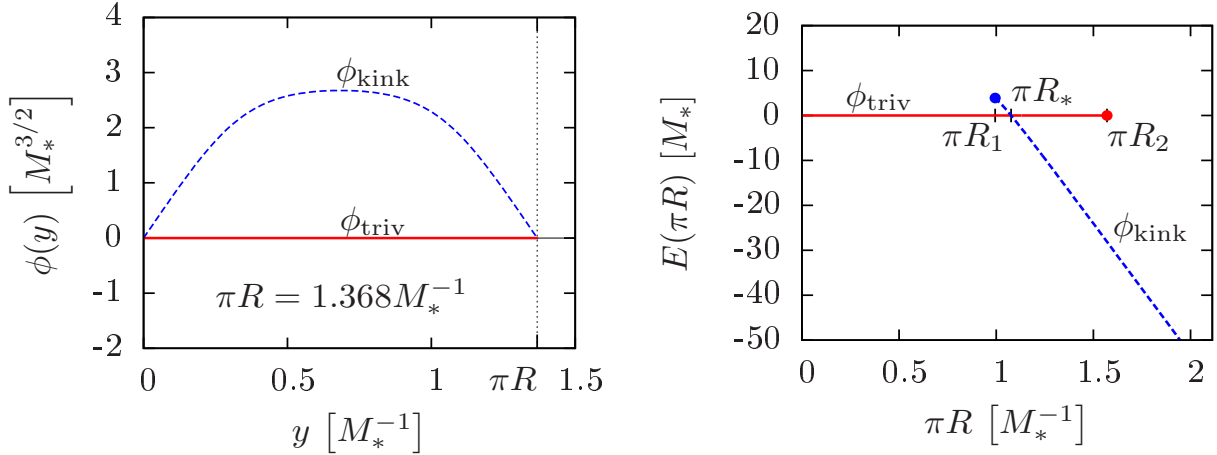}
  \vspace{0.2cm}
  \caption{Profiles (left panel) in the extra dimension interval
    $[0,\pi R]$ of the two possible stable static
  configurations of the Dirichlet scalar field $\phi$, defined by the scalar
  potential
  $V(\phi)=-\frac{1}{2}|\mu^2|\phi^2-\frac{1}{4}|\lambda|\phi^4+\frac{1}{6}|\xi|\phi^6$
  (with $\mu^2=4\ms^2$, $\lambda=4\ms^{-1}$,  $\xi=0.6\ms^{-4}$, and
  $\pi R=1.368\ms^{-1}$).  In the right panel, we show the energy of
  the two stable solutions as a function of $\pi R$, and it is seen how
  the absolute stability of coexisting static configurations is
  determined by the size of the radius $R$. The dots show critical
  points where the scalar perturbations contain a massless mode.} 
  \label{phasediagram}
\end{figure}

In Fig.~\ref{phasediagram} we show an example of a simple setup
defined by the scalar field potential 
$V(\phi)=-\frac{1}{2}|\mu^2|\phi^2-\frac{1}{4}|\lambda|\phi^4+\frac{1}{6}|\xi|\phi^6$,
with $\mu^2=4\ms^2$, $\lambda=4\ms^{-1}$ and $\xi=0.6\ms^{-4}$. In the right panel,
the energy density of two static solutions is plotted as a function of
$R$, showing clearly that below a critical radius $R_{1}$ only the
trivial solution is possible and above a critical radius $R_{2}$ only
the kink solution is possible. For $R_{1}<R<R_{2}$, both solutions are perturbatively stable. At the radius $R_{*}$ the two solutions are 
degenerate, marking the transition from one true vacuum to another
($\phi_{\mathrm{triv}}$ for $R<R_{*}$ and $\phi_{\mathrm{kink}}$ for
$R>R_{*}$). From this we see that the inverse length scale $1/R$ plays
the role of an order parameter of a phase transition, much like
temperature $T$ in finite temperature field theory. For a very small
radius $R$ (analogous to high $T$) the system is stable only around
its trivial solution, with all symmetries restored. As the radius
increases (analogous to $T$ decreasing) the system can undergo a phase
transition, which could be of either first or second order. The analogy with
temperature, however, is not meant to be taken literally. For whereas
the temperature in any 4D effective cosmology must be monotonically
decreasing for most of its history, the orbifold radius $R$ could in
principle increase, decrease or oscillate on very long time scales,
depending on the dynamics of the stabilization mechanism (which we have so far
ignored). 


\section{Kinks on a Warped Background}
\label{warpedxd_fixed}

We now extend previous investigations to the case of a scalar field in
a warped extra dimension, while neglecting any backreaction on the warping from
the scalar field itself. In this case one includes the
effects of the curved metric along the extra dimension on the scalar
field solutions while still ignoring the dynamics of the gravitational
sector. We therefore consider the action 
\be
   S = \int d^5x\sqrt{-g}\,
   \left[\frac{1}{2}g^{MN}(\partial_M \phi)\,\partial_N \phi 
   - V(\phi)\right] \ ,
   \label{action_warped_metric}
\ee
where the form of the metric is now taken to be
\be
   ds^2 = e^{-2\sigma(y)}\gamma_{\mn}(x)dx^\mu dx^\nu - dy^2\ ,
   \label{warped_metric}
\ee
and where $\sigma(y)$ is the
warp-factor and $\gamma_{\mn}$ the 4D metric on slices of constant
$y$. The purpose of considering scalar field configurations on a fixed
background is to explore whether our previous 
results continue to hold in the presence of a warped background in a
regime where we still have semi-analytical control over the
solutions. We postpone a discussion of the full dynamical problem,
including the backreaction on the metric due to the presence of the scalar
field, until the next section. 

\subsection{Kink Scalar in an $AdS_5$ Background}

In the original Randall-Sundrum (RS) model~\cite{Randall:1999ee}, the metric takes
the form (\ref{warped_metric}) with $\sigma(y)=k|y|$ and
$\gamma_{\mn}=\eta_{\mn}$, where $k$ has dimensions of mass and is related to the 5D cosmological
constant of $AdS_5$.  In this background any static nontrivial field
configurations $\bphi(y)$ are solutions of 
\bea
   \bphi'' - 4 k\bphi' 
   - \left.\frac{\partial V}{\partial\phi}\right|_{\bphi} = 0 \ .
   \label{kinkeqads}
\eea
Scalar perturbations around this kink background, $\pphi(x,y)=\phi(x,y)-\bphi(y)$, can be decomposed as
\be  
  \pphi(x,y) = \Sum_n\pphi_{x}^{(n)}(x)\pphi_{y}^{(n)}(y)
  \label{pphidecomp}
\ee
such that the normal modes $\pphi_{x}^{(n)}(x)$ and $\pphi_{y}^{(n)}(y)$ are solutions of
\bea
   {}^{(4)}\Box \pphi_{x} + m_{n}^2\pphi_{x} &=& 0 \\
   \pphi_{y}'' - 4k\pphi_{y}' - (\mu^2(y) - m_{n}^2 e^{2ky})\pphi_{y} &=& 0,
   \label{pphieq}
\eea
where $\mu^2\equiv\left.\frac{\partial^2
  V}{\partial\phi^2}\right|_{\bphi}$ and
${}^{(4)}\Box\equiv\eta^{\mn}\partial_{\mu}\partial_{\nu}$. Taking the
derivative of the kink equation (\ref{kinkeqads}) gives 
\be 
   \pphi_{M}'' - 4k\pphi_{M}' - \mu^2(y)\pphi_{M} = 0 \ ,
\ee
where we have defined $\pphi_{M}\equiv\bphi'$. Thus $\pphi_{M}$ is a massless
solution ($m_{n}^2=0$) of the perturbation equation~(\ref{pphieq}),
although it satisfies mixed boundary conditions rather than the
Dirichlet boundary conditions imposed on $\pphi_{y}$. 

At this point we are already able to state a new result of this work, which is an
extension of the previous result related to the impossibility of
having stable kink solutions with nodes inside the interval. 
Suppose that $\bphi(y)$ happens to have $\delta$ nodes inside the
interval. We have just shown that $\bphi'\equiv \pphi_{M}$ will solve
the equation for a massless excitation, but with mixed boundary
conditions. Since $\bphi$ has $\delta$ nodes, $\bphi'=\pphi_{M}$
must have $\delta+1$ nodes inside the interval. The following inequalities relating the eigenvalues $\lambda^D_{n}$
for the Dirichlet case and the eigenvalues $\lambda^M_{n}$ for a general
mixed boundary condition case~\cite{zettl} hold from Sturm-Liouville theory
\be 
   \lambda^D_{n} \le \lambda^M_{n+2} \le \lambda^D_{n+2} \ .
\ee
Since we have $\lambda^M_{\delta+1}=0$ (i.e. the eigenvalue of the
solution with $\delta+1$ nodes), we can immediately deduce that the
mass-squared of the lowest excitation of the Dirichlet problem must be
negative since $\lambda^D_{\delta-1} \le \lambda^M_{\delta+1}=0$ with
$\delta \ge 1$. 
\begin{prop} In a warped background on a slice of
$AdS_5$, any static solution to equation~(\ref{kinkeqads}),
with $\delta>0$ nodes inside the interval is always unstable.
\end{prop} 
\noindent However, for nodeless static solutions
(when $\delta=0$) the results for the flat case obtained in
\cite{Toharia:2007xe,Toharia:2007xf} cannot
be extended here. Lacking a general stability condition, we will instead propose a weaker sufficient
stability condition for these and other more generic solutions in the next subsection.

\subsection{Kink Scalar on a General Background}

In a general warped background with metric ansatz~(\ref{warped_metric}) the equation for a static scalar
background configuration is 
\be 
   \bphi'' - 4 \sigma'\bphi' 
   - \left.\frac{\partial V}{\partial\phi}\right|_{\bphi} = 0 \ .
   \label{kinkeqsig}
\ee
In this situation, although we have been unable to extend the stability theorems
found earlier, we are still able to find a general
{\it sufficient}
condition for perturbative stability of the background configurations.  

Small perturbations around the background $\bphi(y)$ may be defined as
in (\ref{pphidecomp}). The spectrum of these perturbations consists of
solutions to the eigenvalue problem  
\bea
   \pphi_{y}(y_{1}) = \pphi_{y}(y_{2}) = 0 \\
   \pphi_{y}'' - 4\sigma'\pphi_{y}' 
   - \left[\mu^2(y) - m^2_{n}e^{2\sigma(y)} \right]\pphi_{y} = 0 \ .
   \label{pphieqsig}
\eea
A useful form of this equation is obtained by performing a change of
variables $\ e^{\sigma(y)} dy=dz$ and defining $\sigma(z)=-\frac{2}{3}\ln{(J(z))}$ and
$W(z)=\mu^2(z)e^{-2\sigma(y(z))}$ to yield
\be 
   \frac{(J\pphi)''}{J\pphi}- \frac{J''}{J} - \left(W(z)-m_{n}^2 \right) = 0 \ .
   \label{Jequation}
\ee

To proceed, we make use of the following integral
inequality~\cite{math}. For any function $f(z)$, such that
$f(a)=f(b)=0$, and with $n$ nodes within the interval $[a,b]$, there exists $\rho\in{\cal R}$ such that 
\be 
   \int_a^b e^{-\rho  f''(z)/f(z)}\ dz \ge (n+1) e \sqrt{\rho \pi} \ .
\ee
Applied to~(\ref{Jequation}), this implies
\be
   e^{\rho m_{n}^2} \int_a^b e^{-\rho  [\frac{J''}{J}  + W(z) ]}\ dz \ge (n+1) e \sqrt{\rho \pi} \ ,
\ee
the logarithm of which yields
\be 
   m_{n}^2 \ge \frac{1}{\rho}\ln[ (n+1) e \sqrt{\rho \pi}] 
   - \frac{1}{\rho}\ln\left( \int_a^b e^{-\rho  [\frac{J''}{J}  + W(z) ]}\ dz \right)
\ee
which is a lower bound for the eigenvalues in terms of the
background quantities $\sigma(z)$ and $\mu(z)$ (which are contained in $J$ and
$W$). In the case of the lowest eigenvalue we have 
\be 
   m_{0}^2 \ge \frac{1}{\rho}\ln( e \sqrt{\rho \pi}) 
   - \frac{1}{\rho}\ln\left( \int_a^b e^{-\rho [\frac{J''}{J}  + W(z)]}\ dz \right)
\ee
and so a sufficient condition for perturbative stability ($m_{0}^2\ge 0$) is 
\be 
   \int_a^b e^{-\rho  [\frac{J''}{J}  + W(z) ]}\ dz \le e \sqrt{\rho \pi} \ .
\ee
We may formulate this explicitly in terms of the warp factor
$\sigma(z)$ so that finally, a static solution $\bphi(z)$ of~(\ref{kinkeqsig}), obeying
Dirichlet boundary conditions, is stable if
\be 
   \int_a^b e^{-\rho[-\frac{3}{2}\sigma'' + \frac{9}{4}\sigma'^2  +\mu^2(z)e^{-2\sigma}]}\ dz 
   \le e \sqrt{\rho \pi} \ ,
   \label{suffcond}
\ee
where $\left.\mu^2(z)\equiv\frac{\partial^2 V}{\partial
 \phi^2}\right|_{\bphi_(z)}$ and where the actual value of $\rho$ is
that which extremizes the right hand side. This is a sufficient
condition, but not a necessary one. In order to demonstrate how
effective this {\it weaker} stability  condition can be, we now turn
to a simple example in which the condition can actually be evaluated.  

Consider a flat metric, where $\sigma(y)\equiv 0$, and a trivial
background scalar configuration, i.e., $\bphi(y)\equiv 0$, but where
the 5D scalar potential is allowed to have a tachyonic mass. In this
case equation (\ref{pphieqsig}) becomes 
\be 
   \pphi_{y}'' - \left(\mu^2-m_{n}^2 \right) \pphi_{y}=0 \ .
\ee
If $\pphi_{y}$ has Dirichlet boundary conditions, the solutions to this problem are
\be 
   \pphi_{y}= \sin(\sqrt{m_{n}^2-\mu^2}\ y)
\ee
where $m_{n}^2-\mu^2=n^2\pi^2/L^2$ and $L=b-a$ is the size of the extra
dimension. The mass of the lightest mode is
$m_{0}^2=\mu^2+\pi^2/L^2 $ and so the condition for stability is
$\ m_{0}\le 0\ $, which means that the bulk scalar mass $\mu^2$ can be
negative, but not arbitrarily so: 
\be 
   \mu^2\ge -\pi^2/L^2\ .
   \label{flatexact}
\ee
Therefore in this case (where $\sigma'=\sigma''=0$), our {\it sufficient} condition~(\ref{suffcond}) becomes
\be 
   e^{-\rho  \mu^2} \int_a^b dz \le e \sqrt{\rho \pi} \ ,
\ee
which leads to
\be 
   \mu^2 \ge \frac{1}{\rho}\ln(\frac{L}{e\sqrt{\rho \pi}}) \ .
\ee
The value of $\rho$ that extremizes this bound is $\rho=\pi L^2/e$, and so our weaker bound is
\be 
   \mu^2 \ge - \frac{e\pi}{2} \frac{1}{L^2} \ .
\ee
This result is a factor of $2\pi/e$ weaker than the exact bound~(\ref{flatexact}). 
Nevertheless this result is nontrivial as it clearly demonstrates that it is possible to
have negative bulk masses and retain a stable system.\footnote{The
  stability conditions of the trivial vacuum in the presence of
  negative bulk mass terms in an extra-dimensional scalar field theory
  have been analyzed and generalized to general warped backgrounds in
  \cite{Toharia:2008ug}.}   


\section{Kinks in Gravitating Warped Extra Dimensions}
\label{warpedxd_dynamic}

So far we have examined static scalar field configurations in a fixed
background. We have found that some of the results that were shown to
hold in a flat extra-dimensional background continue to hold in a
fixed warped background, and we have found useful generalizations of other results. 
We now want to include the dynamics of the
gravitational sector and explore how these results can be extended
when the gravitational backreaction is included. Therefore we now seek nontrivial
static field configurations in which the warp factor has its own dynamics
determined by the 5D Einstein equations. 

As soon as we include a dynamical gravitational sector, we are required to
worry about stabilization of the extra dimension. In the above
discussion we assumed that the extra dimension was stabilized and that the
dynamics of the stabilization mechanism were frozen
out. Here we want to include the backreaction of any matter fields on
the 5D metric, and so we must include the dynamics of stabilization.  A natural
question to ask is whether the kink fields of interest
could provide a stabilization mechanism.  Unfortunately, in
\cite{Lesgourgues:2003mi} it was shown that when one considers static
solutions for both the warp factor and a single scalar field, the
lightest scalar perturbative mode (the radion) will be tachyonic
whenever the derivative of the scalar profile vanishes inside the
interval.  In other words, the system is unstable whenever the scalar
field profile passes through an extremum in the bulk. This means that if we insist
on obtaining a nontrivial configuration for a single scalar field with
Dirichlet boundary conditions, we are guaranteed to obtain a tachyonic
radion and the extra dimension will be unstable. To address this issue
we will add extra scalar fields whose purpose will be to stabilize the
radion as in~\cite{Goldberger:1999uk}. 

The resulting system becomes considerably more difficult to analyze
than the case with only one bulk scalar field, particularly with regard
to questions about stability. On the other hand, the case with three
or more scalar fields is formally no more difficult to analyze than
the case with only two scalar fields. Hence we will keep our treatment
general to include an arbitrary number of scalar fields $\chi_{a}$
($a,b=1,\ldots,\N$), although when we consider particular
examples below, we will specialize to the case with only two scalar
fields (a kink field and a non-kink field). For simplicity we will assume throughout that the scalar
fields are only coupled gravitationally. 

We therefore consider the 5D action for gravity and $\N$ free scalar fields
\bea
   S &=& -\frac{\ms^3}{2}\int d^5x\sqrt{-g}\,\left[\R - 2\Lambda\right] \nn
\\
   &+& 
\int d^5x\sqrt{-g}\,
   \left[\Sum_{a=1}^{\N}\frac{1}{2}\,g^{MN}(\partial_M\chi_{a})(\partial_N\chi_{a}) 
   - W(\chi_{a}) -
   \Sum_{i=1,2}\lambda_{i}(\chi_{a})\delta(y-y_{i})\right] \ ,
\eea
where $\ms\equiv(8\pi G)^{-1/3}$, $G$ is the 5D Newton's constant,
$\R$ is the 5D Ricci scalar, and $\Lambda$ is the 5D cosmological
constant. The full scalar potential in the bulk is $W(\chi_{a})$, and
the brane potentials are $\lambda_{i}(\chi_{a})$. As before, we take
the 5D line element of the form 
\be
   ds^2 = e^{-2\sigma(y)}\gamma_{\mn}(x) dx^\mu dx^\nu - dy^2\ ,  
\ee
where $\gamma_{\mn}$ is the induced metric on the 4D hypersurfaces of
constant $y$, which foliate the extra dimension. The 5D
Einstein and field equations are  
\bea
   \sigma'' - \sigma'^2 + \frac{\Lambda}{6} 
   &=& \frac{1}{2\ms^3}\left( \Sum_{a}^{\N} \frac{1}{2}\chi_{a}'^2 
   + \frac{1}{3}\,W(\chi_{a})
   + \frac{2}{3}\Sum_{i=1,2}\lambda_{i}(\chi_{a})\delta(y-y_{i})\right)
   \label{bkg1}
\\
   \sigma'^2 &-& \frac{\Lambda}{6} + \frac{{}^{(4)}\R\,\,}{12}e^{2\sigma} 
   = \frac{1}{6\ms^3} \left( \Sum_{a}^{\N} \frac{1}{2}\chi_{a}'^2 
   - W(\chi_{a})\right)
   \label{bkg2}
\\
   \chi_{a}'' &-& 4\sigma'\chi_{a}' - \frac{\partial W}{\partial\chi_{a}} 
   - \Sum_{i=1,2}\frac{\partial\lambda_{i}}{\partial\chi_{a}}\delta(y-y_{i}) 
   = 0 \ ,
   \label{bkg3}
\eea
where ${}^{(4)}\R$ is the 4D Ricci scalar associated with the induced
4D metric $\gamma_{\mn}$, which we have left arbitrary. The boundary
conditions for the system are determined by Israel junction
conditions at each brane. These are obtained by integrating the
equations of motion over an infinitesimally small interval across each brane, giving 
\bea
   \left[\sigma'\right]_{y_{i}} &\equiv& 
   \Lim_{\epsilon\to\,0} \left[\sigma'(y_{i}+\epsilon) - \sigma'(y_{i}-\epsilon)\right]
   = \frac{1}{3\ms^3}
\left.\lambda_{i}(\chi_{a})\right|_{y_{i}}
   \label{jxn1}
\\
   \left[\chi'_{a}\right]_{y_{i}} &\equiv& 
   \Lim_{\epsilon\to\,0} \left[\chi_{a}'(y_{i}+\epsilon) - \chi_{a}'(y_{i}-\epsilon)\right]
   = \left.\frac{\partial\lambda_{i}}{\partial\chi_{a}}\right|_{y_{i}}\ .
   \label{jxn2}
\eea
These yield $\N$ conditions on each brane, which is exactly the number
of data that need to be specified in order for equations (\ref{bkg1})
and (\ref{bkg3}) to form a well-posed problem. 

Note that the above boundary value problem consists of a system of coupled
nonlinear differential equations. Finding solutions analytically for
such a setup is highly unlikely, although it is still possible to
proceed in the opposite direction, i.e. given a particular analytical
solution one can obtain the setup from which it originates. To do so,
one relies on the powerful method of the
superpotential~\cite{Skenderis:1999mm,DeWolfe:1999cp,Csaki:2000zn}, which can be useful even for two
or more scalar fields (see, for example,~\cite{Batell:2008zm} in the context
of soft-wall models). However, even if one solution is constructed in this
way, there is no guarantee that this is the only solution with the same action. We will now describe how to look for all
possible solutions of a given action using a combination of numerical and graphical
techniques.

\subsection{Multiple Solutions}

Whenever there is more than one static solution to the above
boundary value problem {\it with the same action}, we say that multiple solutions exist. In general, the bulk scalar fields can have Dirichlet boundary
conditions, Neumann boundary conditions or more general mixed boundary
conditions. Here we focus on the case where we have one
kink field $\phi$ (obeying Dirichlet boundary conditions), with the remaining $\N-1$ fields $\chi_{a}$ having
Neumann or mixed boundary conditions. When the profiles of these extra fields
are monotonic, they will tend to stabilize the extra dimension, whereas
if their profiles have vanishing derivatives inside the interval, they will tend to destabilize the
extra dimension~\cite{Lesgourgues:2003mi}. Despite this subtlety, we
will generically refer to the non-kink fields as ``stabilization''
fields.

To find solutions we proceed as follows: we specify the Lagrangian in
the bulk and on one of the branes, and we numerically solve an initial
value problem to determine the profiles of the fields along the extra
dimension. Dirichlet boundary conditions are imposed on the kink field
$\phi$ at the initial brane by demanding that it vanish there. For
this to hold, we assume the kink field has a sufficiently heavy brane
mass so that it decouples from the stabilization fields on the
branes. As a result, the kink field disappears from the junction
conditions (\ref{jxn1})-(\ref{jxn2}), which then yield only $\N$
conditions on the initial brane. This leaves $\N+1$ initial conditions
that need to be specified, which we take to be the boundary values for
the derivatives $\phi'$, $\chi_{1}'\ldots\chi_{\N-1}'$, and
$\sigma'$. After solving the initial value problem for a given choice
of initial conditions, we impose  Dirichlet boundary conditions on
$\phi$ at the final boundary by locating the second brane at a point
where the profile of $\phi$ vanishes. In general, the profile will
vanish at several points along the extra dimension, and one may study
kinks with the desired number of nodes by choosing the location of the
second brane accordingly.  Here, as in the flat case, we are primarily
interested in nodeless kink solutions, and we therefore place the second brane
at the first zero of the profile function.     

We now have a solution to a {\it boundary} value problem whose
boundary conditions on the second brane are not yet known.  We
parameterize the brane potential on the second brane
$\lambda_{2}(\chi_{a})$ in terms of $P$ parameters $\alpha_{b}$ (for example,
the brane tension $\Sigma_2$, the brane mass term $m_{a}^2$ of each scalar, the quartic
coupling of each scalar, etc.)
\bea
   &&\lambda_{2}(\chi_{a}) 
   = f(\Sigma_2,m_{1}^2,m_{2}^2,...,m_{\N}^2,...) \ .
   \label{secondbrane}
\eea
Then the junction conditions (\ref{jxn1})-(\ref{jxn2}) at the second
brane ($i=2$) give $\N$ linear equations for the $P$ unknowns
$\alpha_{b}$. By evaluating the fields on the second brane, and
using the parameterization in (\ref{secondbrane}), we then 
invert the $\N$ junction conditions to determine the $\alpha_{b}$.
If this is possible, then the solution to our initial value problem is
also a solution to a corresponding boundary value problem. From this
we see that we must have $P\geq\N$ in order to guarantee that the
field configuration we obtained is the solution to a corresponding
boundary value problem. If $P=\N$, the $\alpha_{b}$ are uniquely
determined, and there is a unique Lagrangian for which the above field
configuration is a solution. On the other hand if $P>\N$, some of the
$\alpha_{b}$ are arbitrary and so there is a family of solutions for
these final-boundary conditions. In that case there is a family of
Lagrangians which yield the obtained field configuration, and one can
proceed by  focusing on one member of this family. If $P<\N$, the
linear system of parameters $\alpha_{b}$ may be overdetermined, in
which case the obtained field configuration is not a solution to any
corresponding boundary value problem. 

We can find additional solutions by changing the initial-boundary conditions
and repeating the above process. Note that by freely varying the field
derivatives ($\phi'$, $\chi_{1}'\ldots\chi_{\N-1}'$) at the initial
brane and determining the remaining quantities from the junction
conditions, it is possible to leave the initial-brane potential unchanged. This
is necessary in order that the action remains unchanged (it is not
sufficient because part of the action is determined by the final-brane
potential). A solution and the resulting final-boundary conditions
(the $\alpha_{b}$) are then found as before. Since each set of
initial shooting values yields a set of $\alpha_{b}$, each $\alpha_{b}$ is a function of the $\N$ initial-boundary
derivatives. Each $\alpha_{b}$ therefore defines an $\N$-dimensional
surface whose level-surfaces can be projected onto the
$\phi'(y_{1})$-$\chi_{a}'(y_{1})$ parameter space (which is an
$\N$-dimensional space). In the above construction there are $P$ such
quantities, and so $P$ level-surfaces intersect at every point in this
parameter space, representing one solution for this action. The
question of whether multiple solutions exist for the same action is
equivalent to the question of whether the same $P$ surfaces
simultaneously intersect at more than one point in the parameter
space.   

We will now show how this works in two simple examples. In both
cases, we will consider a kink field $\phi$ in addition to just one
stabilization field $\chi$, with no interaction terms among them in the scalar potential. In
both examples there will be regions of parameter space in which two
distinct static configurations are possible for the same action. 

\subsection{Example 1: Quartic Potential}

In both of the following examples we consider a Lagrangian for two scalar fields
\bea
   \lag_{matter} &=& \frac{1}{2}\,g^{MN}(\partial_M\phi)\partial_N\phi 
   - V(\phi) - \Sum_{i=1,2}\beta_{i}(\phi)\delta(y-y_{i}) \nn
\\
   && \mbox{} + \frac{1}{2}\,g^{MN}(\partial_M\chi)\partial_N\chi 
   - U(\chi) - \Sum_{i=1,2}\lambda_{i}(\chi)\delta(y-y_{i}) \ ,
\eea
where $\phi$ is the kink field and $\chi$ is the stabilization field with potentials
\bea
   U(\chi) &=& \frac{1}{2}\,m_{\chi}^2\chi^2
   \label{gw_bulk_pot}
\\
   \lambda_{i}(\chi) &=& \ms^{-1}\left(\frac{1}{2}\mu_{i}^2\chi^2 + \Sigma_{i}\right)\ .
   \label{gw_brane_pot}
\eea
The fact that the second brane potential for $\chi$ is parameterized
in terms of two parameters, $\mu_{2}^2$ and $\Sigma_{2}$, will allow us
to find unique solutions to the boundary conditions on the second
brane. The junction conditions (\ref{jxn1})-(\ref{jxn2}) become 
\bea
   \sigma'(y_{i}) 
   &=& (-1)^{i-1}\frac{1}{6\ms^4}
   \left(\frac{1}{2}\mu_{i}^2\chi^2(y_{i}) + \Sigma_{i}\right)
   \label{example_jxn1}
\\
   \chi'(y_{i}) 
   &=& (-1)^{i-1}\frac{1}{2}\mu_{i}^2\chi(y_{i})\ .
   \label{example_jxn2}
\eea
On the second brane ($i=2$) these can be inverted to give
\bea
   \mu_{2}^2 &=& -2\frac{\chi'(y_{2})}{\chi(y_{2})}
   \label{mu}
\\
   \Sigma_{2} &=& -6\ms^4\sigma(y_{2}) + \chi'(y_{2})\chi(y_{2})
   \label{Sigma}
\eea
so that once we determine the fields on the second brane, we can
extract the boundary conditions (and therefore Lagrangian) to which
those fields are a solution.  

\begin{figure}[t!]
   \includegraphics[width=16cm,height=12cm]{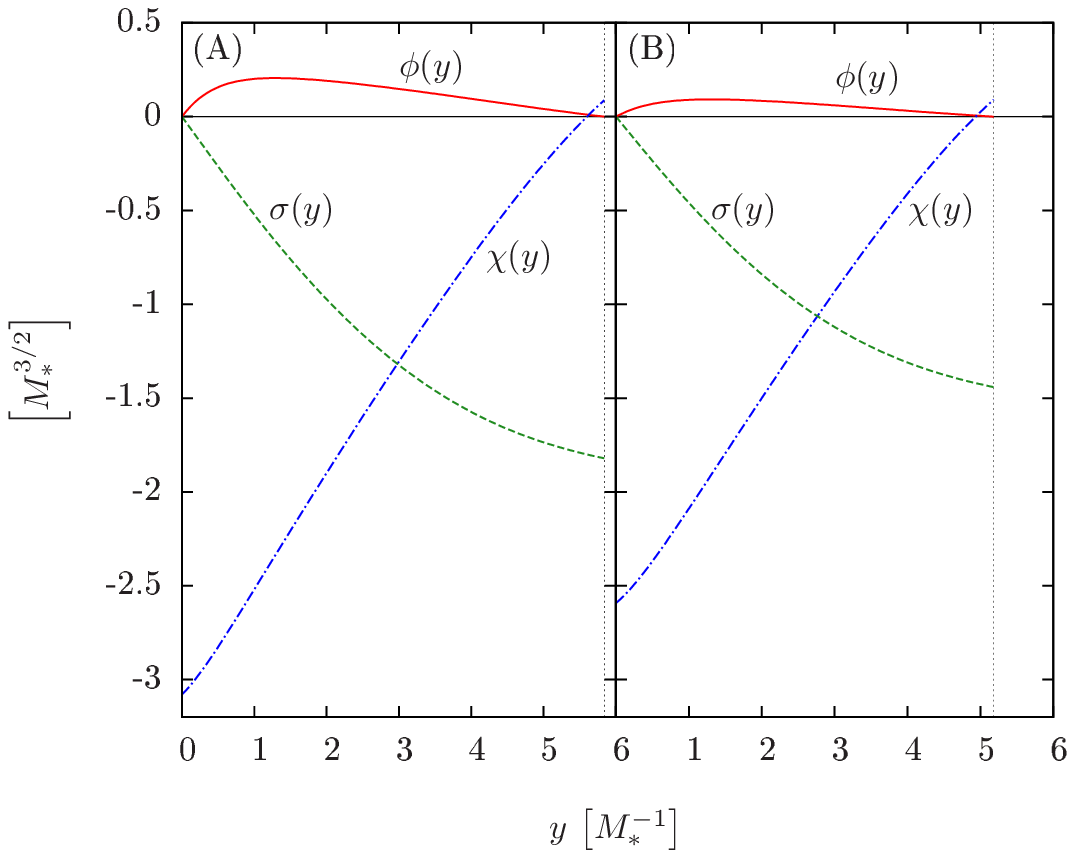}
   \vspace{0.2cm}
   \caption{Profiles of the scalar backgrounds $\phi(y)$ and $\chi(y)$
     as well as the warp factor $\sigma(y)$, showing the two possible solutions
     (panels A and B) to the same boundary value problem defined by the physical parameters
     $m_{\chi}^2=-0.5\ms^2$, $\mu_{1}^2=-0.25\ms^2$,
     $\mu_{2}^2=-8\ms^2$, $\Sigma_{1}=-2\ms^4$,
     $\Sigma_{2}=0.52\ms^4$, $m_{\phi}^2=0.5\ms^2$, $\lambda=2\ms^{-1}$,
     and $\Lambda=0$.  
     } 
   \label{bsoln1}
\end{figure}

\begin{figure}[t!]
   \includegraphics[width=16cm,height=12cm]{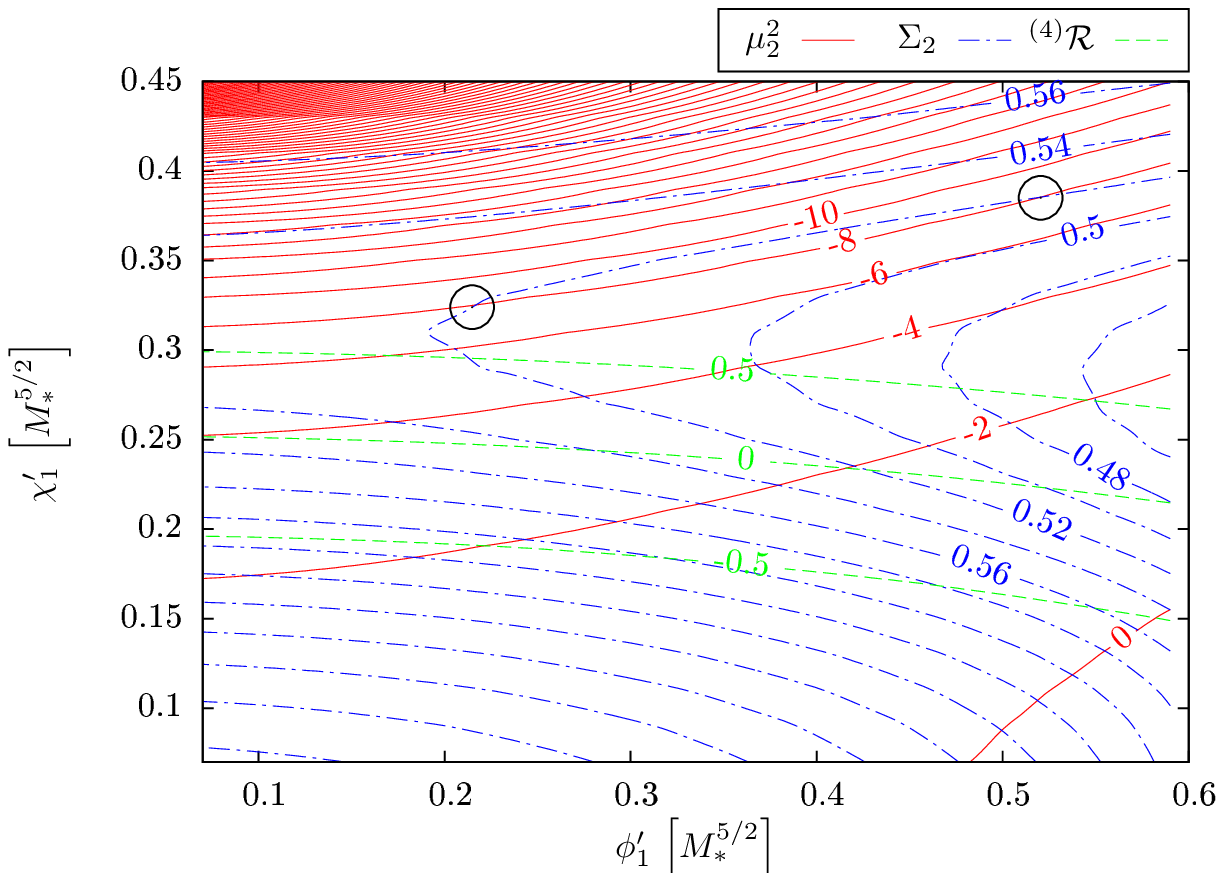}
   \vspace{0.2cm}
   \caption{Level-curves of $\mu_{2}^2$ and $\Sigma_{2}$ in the
     $\phi'_{1}$-$\chi'_{1}$ parameter space for example 1 with
     $m_{\chi}^2=-0.5\ms^2$, $\mu_{1}^2=-0.25\ms^2$, 
     $\Sigma_{1}=-2\ms^4$, $m_{\phi}^2=0.5\ms^2$,
     $\lambda=2\ms^{-1}$, and $\Lambda=0$. Circled are two points in
     the $\phi'_{1}$-$\chi'_{1}$ parameter space with the same values
     of $\mu_{2}^2$ and $\Sigma_{2}$, corresponding to two solutions
     with the same Lagrangian (plotted in Fig.~\ref{bsoln1}).}  
   \label{contour1}
\end{figure}

The only things left to specify are the bulk potential for the kink
field and the initial-boundary conditions. In this first example, we
take the kink potential to be 
\be
   V(\phi) = -\frac{1}{2}\,m_{\phi}^2\phi^2 + \frac{1}{4}\lambda\phi^4 \ .
   \label{kink_pot1}
\ee
Taking the initial brane to be located at $y=0$, we find solutions to
the initial-boundary value problem at this brane with Dirichlet
boundary conditions imposed on the field $\phi$. Examples of nodeless
solutions to the initial value problem are shown in 
Fig.~\ref{bsoln1}. To ensure that $\phi$ obeys Dirichlet boundary
conditions on the second brane, we locate the second brane at the
first point (other than $y=0$) where the profile of $\phi$ vanishes
(the vertical dashed lines in Fig.~\ref{bsoln1}).~\footnote{For certain initial
  conditions, the profile of $\phi$ will blow up before it vanishes
  for a second time. When this happens the initial conditions used do
  not lead to a solution of our boundary value problem.} Once the
position of the second brane is identified, the final-boundary
conditions are determined from (\ref{mu}) and (\ref{Sigma}). By
varying the initial shooting conditions, $\phi_{1}'\equiv\phi'(y_{1})$ and $\chi_{1}'\equiv\chi'(y_{1})$,
and repeating this process of finding solutions, identifying the
location of the second brane, and determining the final-boundary
conditions, we generate level-curves of $\mu_{2}^2$ and
$\Sigma_{2}$. These are plotted in Fig.~\ref{contour1}. Notice that most
$\mu_{2}^2$ contours cross each $\Sigma_{2}$ just once, signifying
that there is a single solution for the corresponding action with the kink potential of
Eq.~(\ref{kink_pot1}). However, some contours cross each other more
than once (see, for example, the circles in Fig.~\ref{contour1}.)  Furthermore there is only a
finite region in the 
$\phi_{1}'$-$\chi_{1}'$ parameter space where solutions exist. If either
$|\phi_{1}'|$ or $|\chi_{1}'|$ are increased sufficiently, the solution to the
initial value problem blows up.
In that case the {\it boundary} value problem has no solution, since a second boundary
where $\phi=0$ does not exist. It is therefore possible to scan
the entire allowed $\phi_{1}'$-$\chi_{1}'$ space and examine whether multiple
solutions with the same action exist. 

\subsection{Example 2: Higher-Order Potential}

In this second example we take a slightly more complicated kink potential
\be
   V(\phi) = -\frac{1}{2}m_\phi^2\phi^2 - \frac{1}{4}\lambda\phi^4 
   + \frac{1}{6}\xi\phi^6\ .
\ee
The other potentials and boundary conditions are the same as in the
previous example, the only difference being the dynamical evolution
of the system due to the new potential $V(\phi)$. We choose this potential because, contrary to the
potential in our first example, in the limit of weak
gravity and flat spacetime, it leads to multiple solutions to the
same boundary value problem~\cite{Toharia:2007xe,Toharia:2007xf} due to the nonlinear nature
of the equations. 

\begin{figure}[t!]
  \includegraphics[width=16cm,height=12cm]{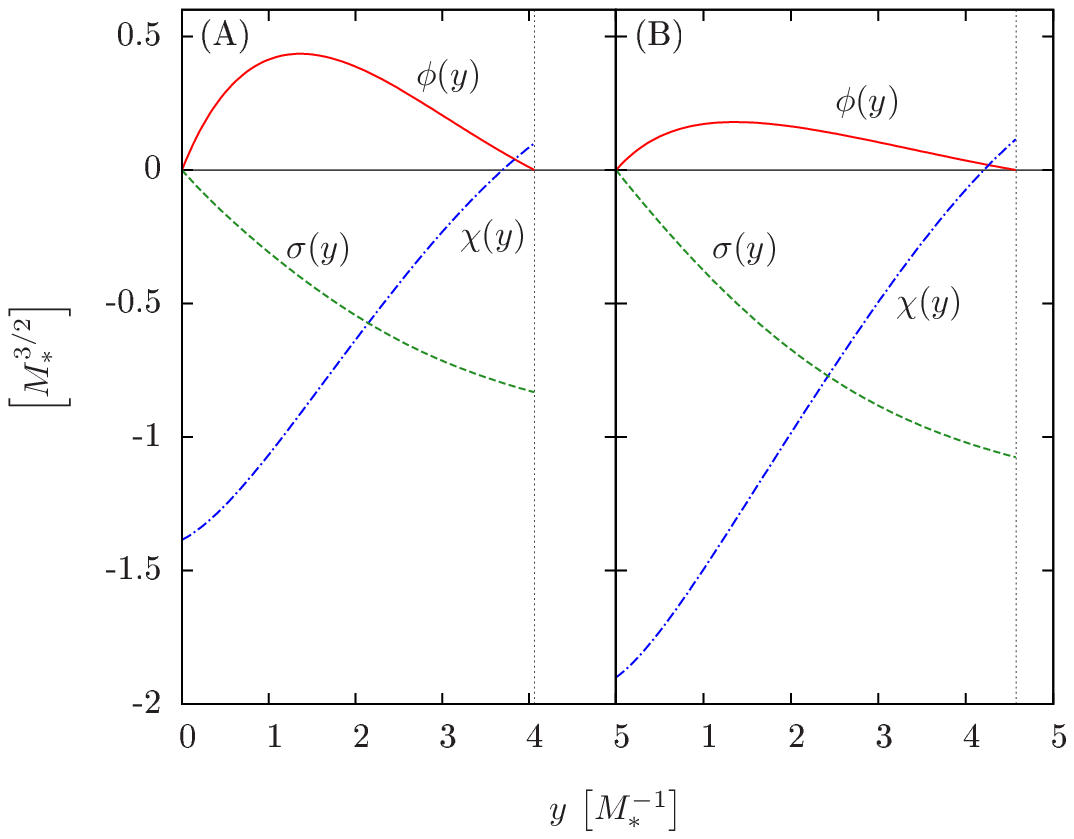}
  \vspace{0.2cm}
  \caption{Profiles of the scalar backgrounds $\phi(y)$ and $\chi(y)$
     as well as the warp factor $\sigma(y)$, showing the two possible solutions
     (panels A and B) to the same boundary value problem defined by
     the physical parameters
$m_{\chi}^2=-0.5\ms^2$,
    $\mu_{1}^2=-0.25\ms^2$, $\mu_2^2=-5\ms^2$, $\Sigma_{1}=-2\ms^4$, $\Sigma_2=0.56\ms^4$, $m_{\phi}^2=0.5\ms^2$,
     $\lambda=2\ms^{-1}$, $\xi=6\ms^{-4}$, and $\Lambda=0$.}
  \label{bsoln2}
\end{figure}

\begin{figure}[t!]
  \includegraphics[width=16cm,height=12cm]{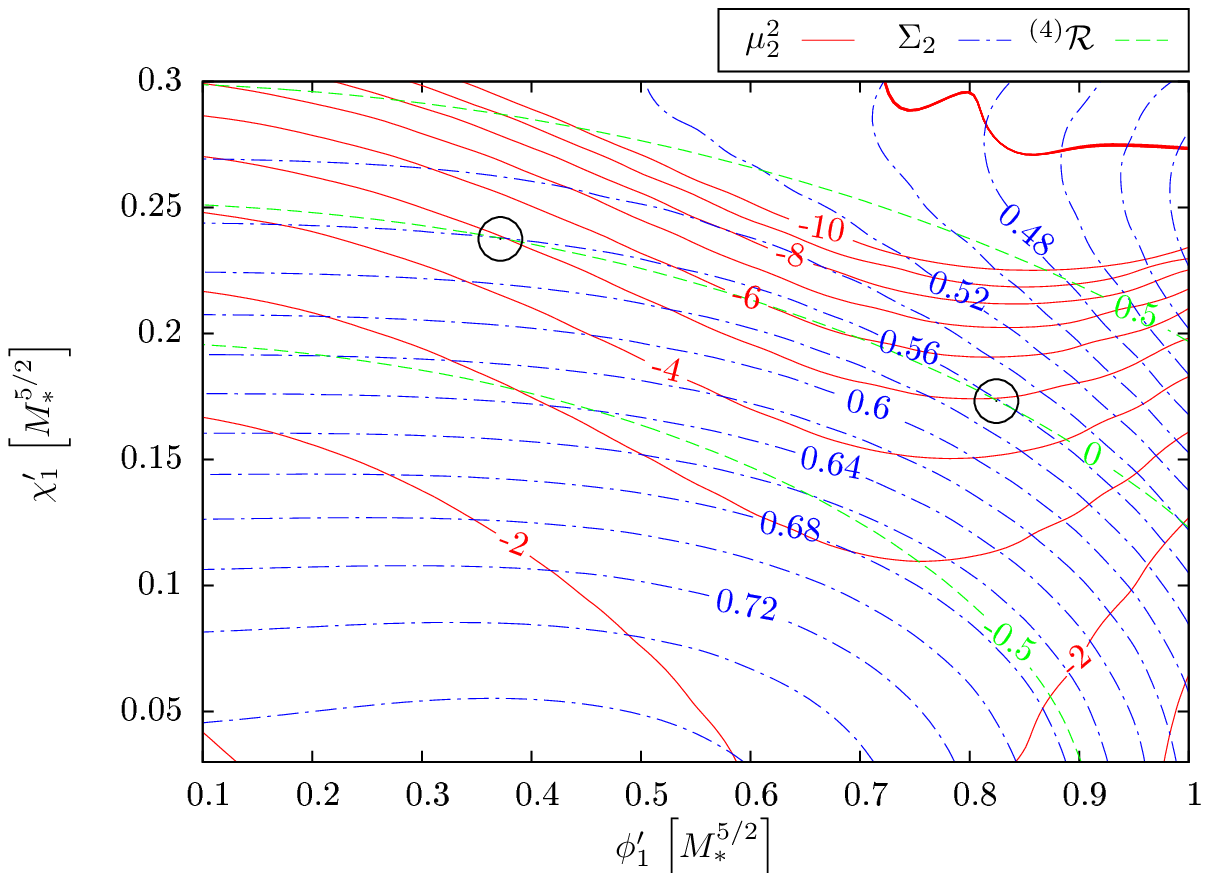}
  \vspace{0.2cm}
  \caption{Level-curves of $\mu_{2}^2$ and $\Sigma_{2}$ in the
    $\phi'$-$\chi'$ parameter space for example 2 with
    $m_{\chi}^2=-0.5\ms^2$, $\mu_{1}^2=-0.25\ms^2$, 
     $\Sigma_{1}=-2\ms^4$, $m_{\phi}^2=0.5\ms^2$,
     $\lambda=2\ms^{-1}$, $\xi=6\ms^{-4}$ and $\Lambda=0$. Circled are
    two points in the $\phi'_{1}$-$\chi'_{1}$ parameter space with the
    same values of $\mu_{2}^2$ and $\Sigma_{2}$, corresponding to two
    solutions with the same Lagrangian. These solutions are plotted in
    Fig.~\ref{bsoln2}.} 
  \label{contour2}
\end{figure}

In our more general setting, including gravity and a stabilization field, we find numerically 
that there exists more than one solution for the same Lagrangian in a large portion of the
parameters space. In Fig.~\ref{bsoln2} we show two such solutions.   
Note that these solutions would be extremely difficult to discover by randomly
guessing initial-boundary conditions. To be more methodical we follow 
the same procedure as before to find level-curves of the
final-boundary conditions, shown in Fig.~\ref{contour2}. Again,
solutions to a particular action will be given by
the intersection of the appropriate contours for the brane mass squared
$\mu_{2}^2$ and brane tension $\Sigma_{2}$. As can be seen, there are
regions in which some contours intersect at more than one point,
showing that multiple solutions for the same action
are possible as expected. In particular, we again circle two such points,
corresponding to the solutions plotted in Figs.~\ref{bsoln2}. As an
interesting remark, note that both of these particular solutions happen
to lie near the region of parameter space where the 4D cosmological
constant vanishes. 


\section{Stability of Solutions}
\label{warpedxd_stability}

Having shown how different nontrivial static field configurations
exist in warped extra dimensions, the next question to ask is whether
these solutions are stable. As we reviewed in section~\ref{flatxd}, in the case
of flat extra dimensions there exist~\cite{Toharia:2007xf}
techniques for determining the stability of such solutions. Indeed, for certain potentials, in that case perturbative stability can be determined
analytically. Unfortunately, in the case of warped extra dimensions,
the question of stability is complicated by the presence of multiple
scalar fields and their coupled dynamics. Here we begin to study the perturbative stability of these kinked configurations. We
derive the linearized equations and reformulate the problem in terms
of a matrix Sturm-Liouville problem. However, the full analysis
requires matrix Sturm-Liouville methods which we omit and leave
for future work. 

We begin by expanding the metric to first-order. Instead of the
coordinates in~(\ref{warped_metric}), in this section it will be more
convenient to choose coordinates so that the metric takes the form 
\be
   ds^2 = a^2(y)\left(\gamma_{\mn}(x)dx^\mu dx^\nu - dy^2\right) \ .
   \label{conformal_warped_metric}
\ee
Working in the generalized longitudinal gauge (see appendix~\ref{ScalarPerts} for details), we 
introduce scalar perturbations $\Phi$ and $\Psi$ and write the perturbed metric as
\be
   ds^2 = a^2(y)\left[(1+2\Phi(x,y))\gamma_{\mn}(x) dx^\mu dx^\nu 
   - (1+2\Psi(x,y))dy^2\right] \ .
   \label{pert_metric}
\ee
Next we expand the $\N$ scalar fields to first-order in small perturbations $\pchi_{a}(x,y)$
\be
   \chi_{a}(x,y) 
   = \bchi{}_{a}(y) + \pchi_{a}(x,y)\ ,
\ee
and compute the linearized Einstein equations, yielding $\N+1$
dynamical equations for $\N+1$ scalar fields ($\N$ fundamental scalars
and one graviscalar, or radion). Since only $\N$ of these equations are
independent, the Einstein constraint equations are used to
eliminate one of the scalar fields in terms of the others (see
Appendix~\ref{ScalarPerts} for details). The resulting $\N$
independent equations are 
\bea
   {}^{(4)}\Box\Psi - \Psi'' 
   - \left(9\h - 2a^2\frac{1}{\bchi_{\N}'}\left.\frac{\partial W}{\partial\chi_{\N}}\right|_{\bchi}\right)\Psi' 
   &-& \left(12\h^2 + 4\h' - \frac{1}{2}{}^{(4)}\R - 4a^2\h\frac{1}{\bchi_{\N}'}\left.\frac{\partial W}{\partial\chi_{\N}}\right|_{\bchi}\right)\Psi \nn \\
   &=& -\frac{4a^2}{3\ms^3}\Sum_{a=1}^{\N-1}
   \left(\left.\frac{\partial W}{\partial\chi_{a}}\right|_{\bchi} 
   - \frac{\bchi_{a}'}{\bchi_{\N}'}\left.\frac{\partial W}{\partial\chi_{\N}}\right|_{\bchi}\right)\pchi_{a}
   \label{pert_einstein}
\eea
\be 
   {}^{(4)}\Box\pchi_{a} - \pchi_{a}'' - 3\h\pchi_{a}'
   - a^2\left.\frac{\partial^2 W}{\partial\chi_{a}^2}\right|_{\bchi}\pchi_{a}
   = - 3\bchi_{a}'\Psi'
   - 2a^2\left.\frac{\partial W}{\partial\chi_{a}}\right|_{\bchi}\Psi\ ,
   \label{pert_field}
\ee
where $\h\equiv\frac{a'}{a}$, ${}^{(4)}\Box\equiv\gamma^{\mn}\partial_{\mu}\partial_{\nu}$, and in equation (\ref{pert_field}), as
throughout, we have assumed that there are no direct couplings between
the 5D scalar fields in the scalar potential (in
Appendix~\ref{ScalarPerts} we derive the general form of these
equations when couplings between the fields are included). These
dynamical equations can be written more compactly as  
\bea
   \Box\Psi + \D^y_{1}\Psi &=& \D^y_{2}\pchi
   \label{compact1}
\\
   \Box\pchi + \D^y_{3}\pchi &=& \D^y_{4}\Psi\ ,
   \label{compact2}
\eea
where $\pchi$ has suppressed discrete indices which run over the
$\N-1$ fundamental scalar fields, $\Psi$ is the graviscalar, the
$\D^y_{i}$ are $y$-dependent differential operators (i.e., linear
differential operators having $y$-dependent coefficients and acting
only on functions of $y$) also with suppressed discrete indices, and
$\Box\equiv\gamma^{\mn}\partial_{\mu}\partial_{\nu}$ is the 4D wave operator.  

The boundary conditions are determined by integrating the equations of
motion across each brane. Integrating equations (\ref{pert_einstein})
and (\ref{pert_field}), these are found to be  
\bea
   [\Psi']_{y_{i}} 
   - \left.2a^2\frac{1}{\bchi_{\N}'}
   \left.\frac{\partial W}{\partial\chi_{\N}}\right|_{\bchi}\Psi'\right|_{y_{i}} 
   &-& 4\left.a^2\h\frac{1}{\bchi_{\N}'}
   \left.\frac{\partial W}{\partial\chi_{\N}}\right|_{\bchi}\Psi\right|_{y_{i}} 
   \nn \\
   &=& \left.\frac{4a^2}{3\ms^3}\Sum_{a=1}^{\N-1}
   \left(\left.\frac{\partial W}{\partial\chi_{a}}\right|_{\bchi} 
   - \frac{\bchi_{a}'}{\bchi_{\N}'}\left.\frac{\partial W}{\partial\chi_{\N}}\right|_{\bchi}\right)\pchi_{a}\right|_{y_{i}}
   \label{pert_einstein_jxn}
\eea
\be
   [\pchi_{a}']_{y_{i}}
   + \left.a^2\left.\frac{\partial^2 W}{\partial\chi_{a}^2}\right|_{\bchi}
   \pchi_{a}\right|_{y_{i}}  
   = 2\left.a^2
   \left.\frac{\partial W}{\partial\chi_{a}}\right|_{\bchi}\Psi\right|_{y_{i}}\ .
   \label{pert_field_jxn}
\ee
These boundary conditions can be put in the form
\bea
   \Psi'(x,y_{i}) &=& A_{1}(y_{i}) \Psi(x,y_{i}) + A_{2}(y_{i}) \pchi(x,y_{i})
   \label{Psibc}
\\
   \pchi'(x,y_{i}) &=& B_{1}(y_{i}) \Psi(x,y_{i}) + B_{2}(y_{i}) \pchi(x,y_{i})\ ,
   \label{pchibc}
\eea
where $A_{1,2}$ and $B_{1,2}$ are functions of $y_{i}$, defined via~(\ref{pert_field_jxn}), and 
we have used~(\ref{Psibc}) in~(\ref{pert_einstein_jxn}) to obtain~(\ref{pchibc}). 

The plan now is to perform a separation of variables in order to obtain a Sturm-Liouville
eigenvalue problem, and then to analyye this eigenvalue problem to determine
stability of the system. Because the 5D equations of motion of the
scalar perturbations are coupled, the correct separation of variables
ansatz is a coupled one 
\bea
   \Psi^{(n)}(x,y) &=& \Psi_{y}^{(n)}(y)\ \ux^{(n)}(x)
   \label{twist1}
\\
   \pchi^{(n)}(x,y) &=& \pchi_{y}^{(n)}(y)\ \ux^{(n)}(x)\ ,
   \label{twist2}
\eea
where $\ux^{(n)}(x)$ is the $n^{th}$ 4D Kaluza-Klein physical mode and
$\Psi_{y}^{(n)}(y)$ and $\pchi_{y}^{(n)}(y)$ are the wave functions. Plugging this ansatz into equations
(\ref{compact1}) and (\ref{compact2}) leads to the following coupled
equations  
\bea
   \Psi_{y}(y)\Box \ux(x) + \ux(x)\D^y_{1}\Psi_{y}(y) 
   &=& \ux(x)\D^y_{2}\pchi_{y}(y)  
\\
   \pchi_{y}(y)\Box \ux(x) + \ux(x)\D^y_{3}\pchi_{y}(y) 
   &=& \ux(x)\D^y_{4}\Psi_{y}(y)\ .
\eea
The separation of variables thus yields a 4D wave equation for $\ux(x)$ 
\be 
   \Box\ux(x) + m_{\ux}^2\ux(x) =0
\ee
and a system of two coupled differential equations
\bea
   \D^y_{1} \Psi_{y}(y) - m_{\ux}^2 \Psi_{y}(y) &=& \D^y_{2}\pchi_{y}(y)\label{coupledeq1} \\
   \D^y_{3} \pchi_{y}(y) - m_{\ux}^2 \pchi_{y}(y) &=& \D^y_{4}\Psi_{y}(y)
\label{coupledeq2}
\eea
with boundary conditions for the profiles 
\bea
   \Psi_{y}'(y_{i}) &=& A_{1}(y_{i})\Psi_{y}(y_{i}) + A_{2}(y_{i})\pchi_{y}(y_{i})\\
   \pchi_{y}'(y_{i}) &=& B_{1}(y_{i})\Psi_{y}(y_{i}) + B_{2}(y_{i})\pchi_{y}(y_{i})\ .
\eea
The system of equations (\ref{coupledeq1}) and (\ref{coupledeq2})
constitute an eigenvalue problem. The stability of the static
background around which we have added scalar perturbations
therefore depends on the existence, or absence, of a negative eigenvalue
$m_{\ux}^2$ associated with a solution to eqs.~(\ref{coupledeq1}) and
(\ref{coupledeq2}).

This situation is somewhat unusual, since generally the Kaluza-Klein
eigenvalue problem arising from dimensional reduction consists of a single 
second order differential equation, which can be put in
standard Sturm-Liouville form. Analyzing that Sturm-Liouville
eigenvalue problem is straightforward, since in particular it is known that the
eigenvalues are bounded from below, and that the eigenfunction
corresponding to the smallest eigenvalue has no zeros
within the interval. Therefore, the question of stability in practical
terms becomes the search for a solution to the Kaluza-Klein equation
such that it contains no nodes. Its associated eigenvalue will be the
lightest possible eigenvalue and, if positive, the system will have no classical instabilities.

In the present case, however, the Kaluza-Klein problem is a system of coupled
differential equations. Consequently, matrix Sturm-Liouville
techniques are required. In order to analyze stability further, one
must extend the theory of oscillations and the concept of nodes of solutions to a higher
dimensional problem. Such an analysis, although rather involved, is underway, and will be presented in a
future work.


\section{Discussion and Outlook}

Braneworld theories generally lead to scalar degrees of freedom that
propagate in the extra-dimensional bulk. Understanding the vacuum
structure of these models in the presence of bulk scalar
fields is therefore a prerequisite to fully appreciating their
phenomenological possibilities. Furthermore, bulk scalars may provide a useful way to localize fermions and build braneworld models purely with field theory (e.g., fat branes and soft walls).

In this work, we have studied the vacuum structure of braneworld
models with one warped extra dimension and multiple bulk scalar
fields. In particular we have focused on static configurations along the
extra space coordinate where one of the fields--with Dirichlet
boundary conditions--acquires a nontrivial kink-like profile. To find
these solutions one needs to solve both the Einstein and the scalar
field equations. In the limit of a flat 5D metric
and weak gravity such solutions are known to exist, and the problem of
finding all possible static configurations as well as determining
their perturbative stability has been addressed and
solved~\cite{Toharia:2007xe,Toharia:2007xf}. Here we have built upon
this previous work to determine how warping along the extra dimension
effects the existence and stability of these kink-like solutions. 

When considering a fixed warped background, it was sufficient to look
for nontrivial solutions for a single scalar field. In this case,
neglecting any backreaction of the scalar field on the gravitational
dynamics, we found that such kink-like solutions do indeed exist. As
in the case of a flat extra dimension, we were able to prove that any
kink-like solution with nodes in the bulk is unstable. Thus we have focused on
nodeless kink solutions and the trivial solution. However, in contrast to the flat case, in the presence
of warping we were only able to find a sufficient condition for
determing the stability of these solutions. We were therfore unable to
analytically determine stability for nontrivial solutions in a warped
background, even when that background is fixed (e.g., in
the Randall-Sundrum model with no backreaction). Instead we were forced to determine
stability numerically.

Including the dynamics of the gravitational sector forces the inclusion of additional scalar fields whose purpose is to
stabilize the size of the extra dimension. In that case we were again
able to find nontrivial kink-like configurations, except now for a
coupled multiple-field system. We have described a general graphical
technique to find all possible static configurations of the background
equations with one kink scalar field and an arbitrary number of
additional ``stabilization'' fields. The technique amounts to
generating solution surfaces by varying the shooting parameters needed
to solve the coupled system of equations. This technique also allows
us to look for multiple solutions with the same action. We have
demonstrated how to implement this technique in two simple examples,
where we considered one kink field and one stabilization field in the
presence of gravity. As in the flat case, when the potential for the
kink field is a higher-order polynomial (leading to higher-order
nonlinearity in the field equations), we found that multiple solutions
may exist for the same action. Interestingly, however, we also found
multiple solutions for the same action when the kink potential was a
fourth-order polynomial, which differs from the result obtained in a flat background.  

We have addressed the issue of stability only partially. We have derived
the full 5D perturbative equations, including gravitational
perturbations, for multiple scalar fields in the presence of a warped
extra dimension. The system of equations constitute a matrix
eigenvalue problem, which must be analyzed using an extension of the
usual theorems coming from oscillation theory or Sturm-Liouville
eigenvalue problems. Such techniques exist in the mathematical
literature but due to the complexity of the task, we have left the
numerical analysis of the general case for a later work.


\acknowledgments
The work of M. Trodden and EJW is supported in part by National
Science Foundation grant PHY-0930521, by Department
of Energy grant DE-FG05-95ER40893-A020 and by NASA ATP grant NNX08AH27G.
M. Trodden is also supported by the Fay R. and Eugene L. Langberg Chair.



\appendix
\section{Scalar Perturbations in the Generalized Longitudinal Gauge}
\label{ScalarPerts}

In this appendix we derive the linearized 5D Einstein and field
equations for scalar perturbations in the bulk. We linearize around a background metric
of the form 
\be
   ds^2 = a^2(y)\left(\gamma_{\mn}(x)dx^\mu dx^\nu - dy^2\right) \ .
\ee
The background Einstein and field equations
in these coordinates are 
\bea
   \h' - \frac{\Lambda}{6}a^2 &=& -\frac{\kappa^3}{2}\left(\Sum_{a}\frac{1}{2}\chi_{a}'^2 
   + \frac{1}{3}a^2\,W(\chi_{a}) 
   + \frac{2}{3}a^2\Sum_{i}\lambda_{i}(\chi_{a})\delta(y-y_{i})\right)
   \label{bkgd_einstein1}
\\
   \h^2 &-& \frac{\Lambda}{6}a^2 + \frac{{}^{(4)}\R\,\,}{12}
   = \frac{\kappa^3}{6}\left(\Sum_{a}\frac{1}{2}\chi_{a}'^2 
   - a^2 W(\chi_{a})\right)
   \label{bkgd_einstein2}
\\
   \chi_{a}'' &+& 3\h\chi_{a}' - a^2\frac{\partial W}{\partial\chi_{a}} 
   - a^2\Sum_{i}\frac{\partial\lambda_{i}}{\partial\chi_{a}}\delta(y-y_{i}) 
   = 0\ ,
   \label{bkgd_field1}
\eea
where $\h\equiv \frac{a'}{a}$, ${}^{(4)}\R$ is the 4D Ricci scalar
with respect to the background 4D metric $\gamma_{\mn}$. To
first-order in scalar perturbations, the 5D metric can be written 
\bea
   ds^2 &=& a^2(y)\left[\left\{(1+2\Phi(x,y))\gamma_{\mn}(x)
   + 2E(x,y)_{|\mn}\right\} dx^\mu dx^\nu\right. \nn
\\
   && \quad\quad\quad\quad\quad\left.\mbox{} 
   + 2B(x,y)_{|\mu} dx^\mu dy - (1+2\Psi(x,y))dy^2\right]
\eea
where ${}_|$ indicates a covariant derivative with respect to the 4D
slices of the bulk. Choosing to work in the generalized longitudinal
gauge, we set $B=E=0$, and the linearized metric simplifies to 
\be 
   ds^2 = a^2(y)\left[(1+2\Phi(x,y))\gamma_{\mn}(x)dx^\mu dx^\nu 
   - (1+2\Psi(x,y))dy^2\right]\ .
\ee
We also expand the scalar fields to first-order
\be
   \chi_{a}(x,y) = \bchi_{a}(y) + \pchi_{a}(x,y)\ ,
\ee
where the fields $\bchi_{a}$ obey the background equations of motion
(\ref{bkgd_einstein1})-(\ref{bkgd_field1}) above and $ \pchi_{a}(x,y)$ are small perturbations.  The linearized
Einstein and field equations are  
\be
   2\Phi + \Psi = 0
   \label{pert_einstein1}
\ee
\be
   \Phi' - \h\Psi = -\frac{1}{3\ms^3}\Sum_{a}\bchi_{a}'\pchi_{a}
   \label{pert_einstein2}
\ee
\bea
   {}^{(4)}\Box(2\Phi + \Psi) - 4\Phi'' + 8\h'\Psi 
   &+& 8\h^2\Psi + 4\h(\Psi' - 3\Phi') + \frac{2}{3}{}^{(4)}\R\Phi  \nn \\
   &=& \frac{4}{3\ms^3}\Sum_{a}\left(\bchi_{a}'\pchi_{a}' - \bchi_{a}'^2\Psi 
   + a^2\left.\frac{\partial W}{\partial\chi_{a}}\right|_{\bchi}\pchi_{a}\right)
   \label{pert_einstein3}
\eea
\bea
   {}^{(4)}\Box\Phi - 4\h\Phi' &+& 4\h^2\Psi + \frac{1}{3}{}^{(4)}\R\Psi  \nn \\
   &=& -\frac{1}{3\ms^3}\Sum_{a}\left(\bchi_{a}'\pchi_{a}' - \bchi_{a}'^2\Psi
   - a^2\left.\frac{\partial W}{\partial\chi_{a}}\right|_{\bchi}\pchi_{a}\right)
   \label{pert_einstein4}
\eea
\be 
   {}^{(4)}\Box\pchi_{a} - \pchi_{a}'' - 3\h\pchi_{a}' 
   + a^2\Sum_{b}\left.\frac{\partial^2 W}{\partial\chi_{a}\partial\chi_{b}}\right|_{\bchi}\pchi_{b}
   = -2\bchi_{a}''\Psi - \bchi_{a}'(\Psi' - 4\Phi' + 6\h\Psi)\ ,
   \label{pert_field1}
\ee
where ${}^{(4)}\Box\equiv\gamma^{\mn}\partial_{\mu}\partial_{\nu}$ is the 4D wave operator. Applying the constraint
equation~(\ref{pert_einstein1}) to
equations~(\ref{pert_einstein2})-(\ref{pert_field1}), and making use
of the background equations (\ref{bkgd_einstein1})-(\ref{bkgd_field1}), yields 
\be
   \Psi' + 2\h\Psi
   = \frac{2}{3\ms^3}\Sum_{a}\bchi_{a}'\pchi_{a}
   \label{pert_einstein5}
\ee
\be
   \Psi'' + 5\h\Psi' 
   + \left(4\h' + 4\h^2 - \frac{1}{6}{}^{(4)}\R 
   + \frac{2}{3\ms^3}\Sum_{a}\bphi_{a}'^2\right)\Psi
   = \frac{2}{3\ms^3}\Sum_{a}\left(\bchi_{a}'\pchi_{a}' 
   + a^2\left.\frac{\partial W}{\partial\chi_{a}}\right|_{\bchi}\pchi_{a}\right)  
   \label{pert_einstein6}
\ee
\be
   {}^{(4)}\Box\Psi - 4\h\Psi'
   - \left(8\h^2 - \frac{1}{3}{}^{(4)}\R -\frac{2}{3\ms^3}\Sum_{a}\bchi_{a}'^2\right)\Psi
   = \frac{2}{3\ms^3}\Sum_{a}\left(\bchi_{a}'\pchi_{a}' 
   - a^2\left.\frac{\partial W}{\partial\chi_{a}}\right|_{\bchi}\pchi_{a}\right)
   \label{pert_einstein7}
\ee
\be
   {}^{(4)}\Box\pchi_{a} - \pchi_{a}'' - 3\h\pchi_{a}' 
   + \Sum_{b}\left(
   a^2\left.\frac{\partial^2 W}{\partial\chi_{a}\partial\chi_{b}}\right|_{\bchi}
   - \frac{2}{\ms^3}\bchi_{a}'\bchi_{b}' \right)\pchi_{b}
   = -2\left(3\h\bchi_{a}' 
   - a^2\left.\frac{\partial W}{\partial\chi_{a}}\right|_{\bchi} \right)\Psi\ .  
   \label{pert_field2}
\ee
We obtain a 5D wave-like equation for $\Psi$ by subtracting~(\ref{pert_einstein6}) from (\ref{pert_einstein7}) to give 
\be
   {}^{(4)}\Box\Psi - \Psi'' - 9\h\Psi' 
   - \left(4\h' + 12\h^2 - \frac{1}{2}{}^{(4)}\R\right)\Psi
   = -\frac{4}{3\ms^3}\Sum_{a} 
   a^2\left.\frac{\partial W}{\partial\chi_{a}}\right|_{\bchi}\pchi_{a} \ .
   \label{pert_einstein8}
\ee
Equations~(\ref{pert_einstein8}) and~(\ref{pert_field2}) comprise
$\N+1$ dynamical equations for the $\N+1$ perturbation variables
$\Psi$ and $\pchi_{a}$. However, since these variables are connected
through the constraint (\ref{pert_einstein5}), only $\N$ of them are
independent. Therefore we may use (\ref{pert_einstein5}) to eliminate
one of the variables in terms of the others. Choosing to eliminate the
$\N$th scalar field, $\pchi_{\N}$, in terms of $\pchi_{a<\N}$ and
$\Psi$, equation (\ref{pert_einstein5}) gives  
\be
    \pchi_{\N} = 
   -\frac{1}{\bchi_{\N}'}\left(\Sum_{b=1}^{\N-1}\bchi_{b}'\pchi_{b} 
   - \frac{3\ms^3}{2}(\Psi' + 2\h\Psi) \right)\ .
\ee
(Note that we cannot eliminate $\Psi$ in terms of the scalar fields
$\pchi_{a}$, since this requires an integration over unknown
functions. This is understandable since doing so would amount to
reducing the problem to one in flat spacetime, which ought to be
impossible.) Substituting this into (\ref{pert_einstein8}) and
(\ref{pert_field2}) and rearranging gives  
\bea
   {}^{(4)}\Box\Psi - \Psi'' 
   - \left(9\h - 2a^2\frac{1}{\bchi_{\N}'}\left.\frac{\partial W}{\partial\chi_{\N}}\right|_{\bchi}\right)\Psi' 
   &-& \left(12\h^2 + 4\h' - \frac{1}{2}{}^{(4)}\R -
   4a^2\h\frac{1}{\bchi_{\N}'}\left.\frac{\partial
     W}{\partial\chi_{\N}}\right|_{\bchi}\right)\Psi \nn \\ 
   &=& -\frac{4a^2}{3\ms^3}\Sum_{a=1}^{\N-1}
   \left(\left.\frac{\partial W}{\partial\chi_{a}}\right|_{\bchi} 
   - \frac{\bchi_{a}'}{\bchi_{\N}'}\left.\frac{\partial W}{\partial\chi_{\N}}\right|_{\bchi}\right)\pchi_{a}
   \label{pert_einstein9}
\eea
\bea
   &&{}^{(4)}\Box\pchi_{a} - \pchi_{a}'' - 3\h\pchi_{a}'
   - a^2\Sum_{b=1}^{\N-1}\left(
   \left.\frac{\partial^2 W}{\partial\chi_{b}\partial\chi_{a}}\right|_{\bchi} 
   + \frac{\bchi_{b}'}{\bchi_{\N}'}\left.\frac{\partial^2
     W}{\partial\chi_{\N}\partial\chi_{a}}\right|_{\bchi}\right)\pchi_{b}
   \quad\quad  \nn \\ 
   &&\mbox{} = - 3\left(\bchi_{a}' 
   - \frac{\ms^3a^2}{2\bchi_{\N}'}\left.\frac{\partial^2 W}{\partial\chi_{\N}\partial\chi_{a}}\right|_{\bchi} \right)\Psi'
   - 2a^2\left( 
   \left.\frac{\partial W}{\partial\chi_{a}}\right|_{\bchi}
   + \frac{3\ms^3\h}{2\bchi_{\N}'}\left.\frac{\partial^2 W}{\partial\chi_{\N}\partial\chi_{a}}\right|_{\bchi} \right)\Psi\ .
   \label{pert_field3}
\eea
We may write this more compactly as
\bea
  && {}^{(4)}\Box\Psi + \D_1^y\Psi -
\Sum_{a=1}^{\N-1}(\D_2^y)_{a}\pchi_{a}\ =0
   \label{pert_einstein_compact1}
\\
   && {}^{(4)}\Box\pchi_{a} + \Sum_{b=1}^{\N-1}(\D_3^y)_{ab}\pchi_{b} 
   - (\D_4^y)_{a}\Psi\ =0,
   \label{pert_field_compact1}
\eea
where
\bea
   \D_1^y &\equiv& -\partial_y^2 
   - \left(9\h - 2a^2\frac{1}{\bchi_{\N}'}\left.\frac{\partial W}{\partial\chi_{\N}}\right|_{\bchi}\right)\partial_y 
   - \left(12\h^2 + 4\h' - \frac{1}{2}{}^{(4)}\R -
   4a^2\h\frac{1}{\bchi_{\N}'}\left.\frac{\partial
     W}{\partial\chi_{\N}}\right|_{\bchi}\right)  \nn 
\\
   (\D_2^y)_{a} &\equiv& -\frac{4a^2}{3\ms^3}
   \left(\left.\frac{\partial W}{\partial\chi_{a}}\right|_{\bchi} 
   - \frac{\bchi_{a}'}{\bchi_{\N}'}\left.\frac{\partial W}{\partial\chi_{\N}}\right|_{\bchi}\right)  \nn
\\
   (\D_3^y)_{ab} &\equiv& -\delta_{ab}(\partial_y^2 + 3\h\partial_y) 
   + \M_{ab}  \nn
\\
   \M_{ab} &\equiv& -a^2\left(
   \left.\frac{\partial^2 W}{\partial\chi_{b}\partial\chi_{a}}\right|_{\bchi} 
   + \frac{\bchi_{b}'}{\bchi_{\N}'}\left.\frac{\partial^2 W}{\partial\chi_{\N}\partial\chi_{a}}\right|_{\bchi}\right)  \nn
\\
   (\D_4^y)_{a} &\equiv& -3\left(\bchi_{a}' 
   - \frac{\ms^3a^2}{2\bchi_{\N}'}\left.\frac{\partial^2 W}{\partial\chi_{\N}\partial\chi_{a}}\right|_{\bchi} \right)\partial_y
   - 2a^2\left( 
   \left.\frac{\partial W}{\partial\chi_{a}}\right|_{\bchi}
   + \frac{3\ms^3\h}{2\bchi_{\N}'}\left.\frac{\partial^2 W}{\partial\chi_{\N}\partial\chi_{a}}\right|_{\bchi} \right) \ .
\eea
If we suppress the discrete indices in equations
(\ref{pert_einstein_compact1}) and (\ref{pert_field_compact1}) they
take an even simpler form 
\bea
   {}^{(4)}\Box\Psi + \D_1^y\Psi &=& \D_2^y\pchi
   \label{pert_einstein_compact2}
\\
   {}^{(4)}\Box\pchi + \D_3^y\pchi &=& \D_4^y\Psi\ .
   \label{pert_field_compact2}
\eea
This is the generic form that the scalar perturbation equations of
motion take for $\N-1$ coupled scalar fields and a graviscalar. 


\end{document}